\documentclass[12pt]{iopart}
\usepackage{graphicx,color}
\newcommand{\eqn}[1]{eq.~(\ref{#1})}

\usepackage{graphicx,color}
\usepackage{epstopdf}

\begin{document}
%\draft

\title{ 0.4 and 0.7 conductance anomalies in quantum point contacts}
\author{ A. M. Bychkov$^{1,2}$ and T. M. Stace$^1$\footnote{Present address: Department of Physics, University of Queensland, Brisbane, Australia, stace@physics.uq.edu.au.}}
\address{(1) Department of Applied Mathematics and Theoretical Physics,
University of Cambridge, Wilberforce Road, Cambridge CB3 0WA, UK\\
(2) Microelectronics Research Centre, Cavendish Laboratory,
University of Cambridge, Madingley Road, Cambridge CB3 0HE, UK}
\date{\today}

\begin{abstract}

Self-consistent modelling based on local spin-density formalism is
employed to calculate conductance of quantum point contacts at
finite temperatures. The total electrostatic potential exhibits
spin-dependent splitting, which persists at temperatures up to 0.5
K and gives rise to the anomalies at 0.4 and 0.7 of the
conductance quantum $(2e^2/h)$ occurring simultaneously in the
absence of external magnetic fields.

\end{abstract}
\pacs{73.61.-r, 71.15.Mb, 71.70.Gm}

\maketitle

\section{Introduction}

The quantisation of quantum wire conductance in multiples of
$G_0=2e^2/h$ has long since been observed~\cite{whar,wees} and
understood~\cite{but}. The extreme sensitivity of short quantum
wires formed in the constriction of quantum point contacts (QPCs)
to the local electrostatic environment is demonstrated in their
use as charge detectors capable of monitoring single-electron
tunnelling in real time~\cite{fiel,vand}. In addition, the first
conductance plateau exhibits a series of non-integer features,
most notably at $0.7G_0$~\cite{thom}, observed in various
semiconductor quantum wires under a range of experimental
conditions~\cite{fitz}. It is largely agreed that the non-integer
conductance features are due to electron-electron interactions of
the confined 2DEG, consistent with the fact that the 0.7 structure
evolves into a spin-split subband at $0.5G_0$ with an in-plane
magnetic field~\cite{thom}, but a comprehensive understanding of
the effect is yet to emerge. Proposals based on microscopic
many-body theory ascribe the observed conductance behaviour to the
spontaneous spin polarisation~\cite{ber,bych,hiro,ber2,hav}, and
Kondo-like correlated spin state~\cite{cron, meir} in the QPC,
whereas phenomenological models assume the pinning of one of the
spin subbands to the chemical potential at the opening of a new
conductance channel~\cite{kris,bruu,reil}.

In this paper, we use the local spin-density approximation
(LSDA) model developed by Berggren \textit{et al.}~\cite{ber2}.  The model is modified  to include 
the case of finite temperatures, by replacing the zero-temperature Fermi distribution with the finite temperature version.   We use the LSDA results to calculate the conductance
using a rigorous form of the transmission coefficient obtained
from the self-consistent electrostatic potential. We find that the
transverse modes split into spin-dependent subbands, both above
and below the chemical potential. This results in conductance
features at $0.4G_0$ and $0.7G_0$, consistent with the latest  experimental results of Crook \textit{et al}.\ \cite{crook}.  In our model, the features persist at temperatures up to 0.5
K, and agrees qualitatively with recent results \cite{crook}. %however their temperature dependance does not agree with experimental results.

\section{Methods}

\subsection{Self-consistent electrostatic potential}

\begin{figure*}
\includegraphics[width=15cm]{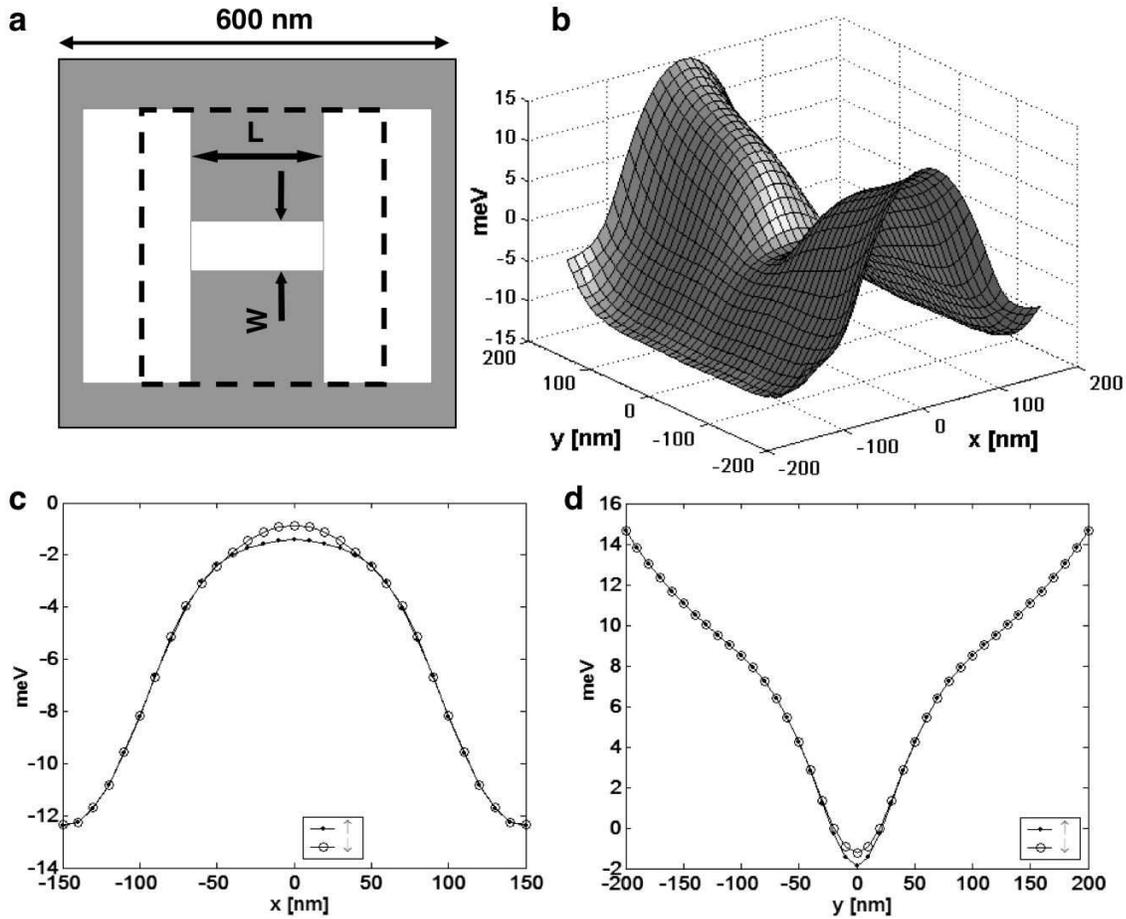}
\caption{(a) Top view of the QPC device showing lithographically
defined openings surrounded by the metallic gate. Dashed line
outlines the region of consideration. (b) Total electrostatic
potential for spin up electrons at $T = 0.1$ K and $V_g=-0.5007$
V. (c,d) Saddle-point longitudinal and transverse, respectively,
cross-sections of the spin up and spin down potentials show spin
gaps in the absence of the magnetic field.} \label{fig:pots}
\end{figure*}

Our simulated QPC device is a modulation-doped heterostructure,
which consists of an AlGaAs spacer, an AlGaAs donor layer
(uniformly doped at $\rho_d=6\times 10^{17}$ cm$^{-3}$), and a
GaAs cap grown epitaxially on a GaAs substrate. A plane metallic
gate is deposited on top of the cap; two $100\times 400$ nm$^2$
rectangular openings connected by a narrow channel 200 nm long and
10 nm wide are then etched through the gate (see
Figure~\ref{fig:pots}a). Applying a negative voltage between the
gate and the substrate creates electrostatic confinement of the
2DEG at the heterojunction 70 nm below the gate plane thus forming
a QPC.  By ensuring the constriction region is far from the boundaries, we may neglect boundary effects, as discussed in \cite{bych}.

To obtain the total spin-dependent electrostatic potentials for
the confined 2DEG, we employ an LSDA theoretical framework
described in detail elsewhere~\cite{ber2,bych,yak,PhD}.  The robustness of this computational method, at zero temperature, has been established in these studies. The model
takes into account electron exchange-correlation effects, whereby
the exchange term $U^{\sigma}_{x}\propto -\sqrt{n^{\sigma}}$ is a
function of the 2D electron density, and the correlation term is
taken in the parameterized form of Tanatar and Ceperley~\cite{tan}
as an interpolation between the fully spin-polarised and
non-polarised cases~\cite{nano}. Single particle LSDA equations
for spin up and spin down electrons are solved self-consistently
via an iterative procedure with the input values for the
potentials derived from the semiclassical Thomas-Fermi
approximation.   For our system, we have a total of around 300 electrons in the region being modelled (yielding an electron density of about $8\times10^{10}\textrm{ cm}^{-2}$), of which around 10 are in the constriction.

Finite temperature is included by calculating the electron density
at each iteration as
\begin{equation}
n^{\sigma}(x,y)=\sum_{i}|\psi_i^{\sigma}(x,y)|^2f(E_i^{\sigma},\mu)
\label{eq:density}
\end{equation}
where $E_i^{\sigma}$ and $\psi_i^{\sigma}(x,y)$ are the
eigenenergies and eigenvectors of the LSDA equations,
respectively, and
$f(E_i^{\sigma},\mu)=(1+e^{(E_i^{\sigma}-\mu)/kT})^{-1}$ is the
Fermi-Dirac distribution function with the chemical potential set
at $\mu =0$.

\subsection{Conductance}

Having computed the effective single particle potential, we can
now calculate the conductance in the linear regime according to
\begin{equation}
G(V_g)=-\frac{G_0}{2}\sum_{\sigma}\int_{-\infty}^{+\infty}
T_{V_g}^{\sigma}(E)\frac{\partial f(E)}{\partial E}
dE\label{eq:cond}
\end{equation}
where $T_{V_g}^{\sigma}(E)$ is the spin and energy dependent
transmission coefficient, and $f(E)$ is the Fermi-Dirac
occupation. The dependence on an external gate voltage, $V_g$, is
included explicitly here, but will be dropped subsequently.  To
compute the transmission coefficients we make a 1D approximation
for the longitudinal potential: we consider a particle scattering
from a 1D potential given by the lowest lateral subband energy,
determined from the self-consistent potential.  The details of the
calculation are described in the Appendix.

\section{Results}

Figure~\ref{fig:pots} shows a typical self-consistent total
electrostatic potential and its longitudinal and transverse
cross-sections at the saddle point. Spin-dependent splitting of
the potentials occurs in the absence of an external magnetic
field, driven mostly by the electron exchange
term~\cite{ber,ber2,bych,star}. In the following, we consider only
the central $300\times 400$ nm$^2$ region of the device where the
potentials are not affected by the finite size of the lithographic
gate openings on either side of the channel.

\begin{figure}
\includegraphics[width=15cm]{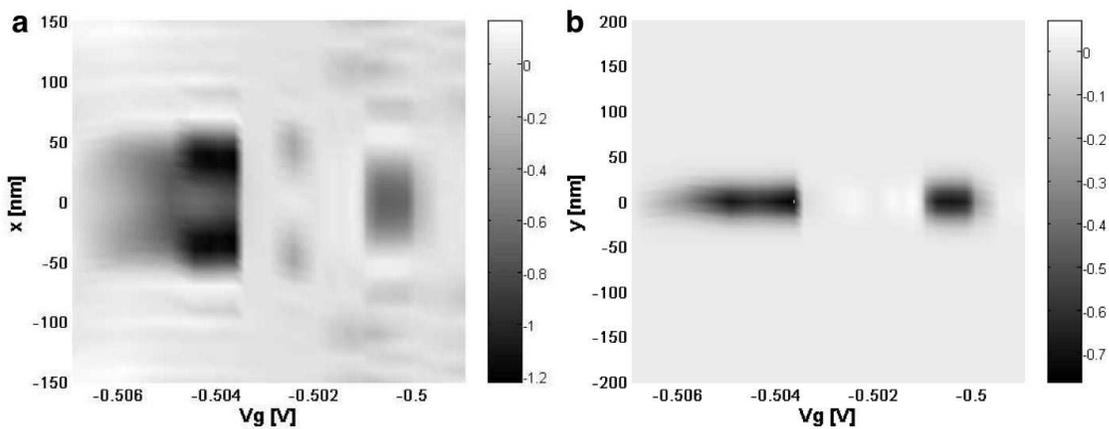}
\caption{Difference between spin up and spin down potentials along
(a) the longitudinal centreline, (b) in the lateral direction
through the middle of the saddle, for different gate voltages at
$T=0.1$ K. The colour bars show the scale in meV.}
\label{fig:potsdiff}
\end{figure}

Figure~\ref{fig:potsdiff} shows the difference between the spin up
and spin down potentials along the $x$ and $y$ directions, passing
through the middle of the saddle. There are clearly two regions in
which the potentials are split: in the region around $V_g=-0.504$
V and $V_g=-0.501$ V. This splitting is the origin of the features
that we find in the conductance. Figure~\ref{fig:potsdiff}b shows
the splitting in the lateral direction, demonstrating that the
splitting is localised around the saddle-point. The constriction
thus forms a quasi-1D channel. We note the applied gate voltages
here differ in scale from experiments due to different device
geometries.

\begin{figure}
\includegraphics[width=8cm]{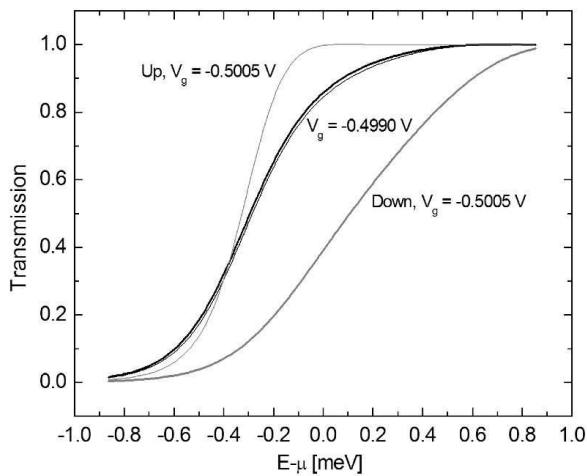}
\caption{Transmission coefficient versus incoming energy at
$T=0.1$ K, at different gate voltages. The heavier lines are spin
down and lighter lines are spin up. Energy is defined with respect
to the chemical potential.  Note that at $V_g=-0.4990 V$ the up and down traces are almost the same.} \label{fig:transm}
\end{figure}

As described above and in the Appendix, we compute the
transmission, $T(E)$ as a function of the incident energy, which
is used in Eq.~(\ref{eq:cond}) to calculate the conductance. Some
representative examples are shown in Figure~\ref{fig:transm} for
two different gate voltages.  At $V_g=-0.501$ V where the
splitting is significant, the transmission coefficients are very
different for the two spin configurations.  At $V_g=-0.4990$ V the
splitting is very small, so the transmission coefficients are very
similar. The width of the step in the transmission is a measure of
the flatness of the top of the tunnel barrier; compare with
Figure~\ref{fig:pots}c, where the  barrier vanishes over an energy
scale of about 2 meV. Since the temperature for this example is
0.1K, the derivative of the Fermi-Dirac distribution in
Eq.~(\ref{eq:cond}) is only $\sim0.01$ meV wide. Over this energy
scale, the transmission coefficients are essentially constant (see
Figure~\ref{fig:transm}), so to a good approximation
\begin{equation}
G=G_{0}[T^{\uparrow}(\mu)+T^{\downarrow}(\mu)]/2. \label{eq:cond2}
\end{equation}
Substituting the numerical values from Figure~\ref{fig:transm} for
$T^{\uparrow}(\mu)$ and $T^{\downarrow}(\mu)$ at $V_{g}=-0.501$
gives $G\approx0.7G_{0}$.

\begin{figure}
\includegraphics[width=8cm]{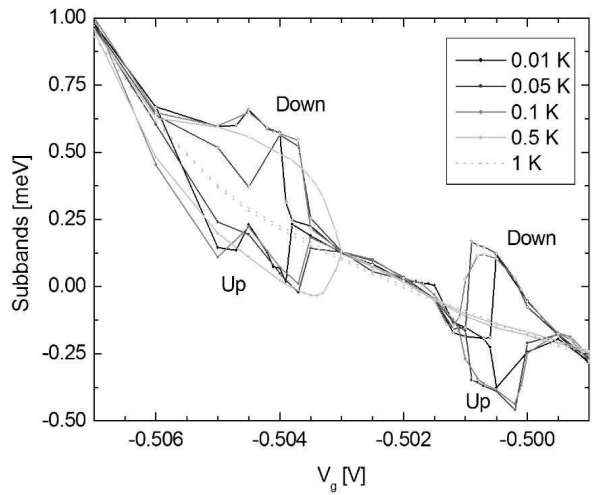}
\caption{Subband energy of the lowest transverse mode for
different gate voltages. The energies are defined with respect to
the chemical potential. 
%Solid points (lower curves) are spin up subbands, and empty points (upper curves) are spin down subbands.
Lower curves are spin up subbands, and upper curves are spin down subbands.  Note that there is almost no splitting at 1K.} 
\label{fig:subbands}
\end{figure}

The variation of the transmission probability with $V_{g}$  can be
most readily understood by considering the dependence of the
subband energies on $V_{g}$, shown in fig.~\ref{fig:subbands}
(right).  The two splittings of the self-consistent potentials
give rise to commensurate splittings in the subband energies.  The
splitting around $V_{g}=-0.504$ V does not track the chemical
potential, but instead, the spin up subband dips below the
chemical potential. This increases the transmission coefficient
for the spin up component, giving rise to a peak in the
conductance. Similarly, the spin down subband rises above the
chemical potential around $V_{g}=-0.501$ V, which suppresses the
transmission coefficient for the spin down component, leading to a
dip in the conductance.

\begin{figure}
\includegraphics[width=8cm]{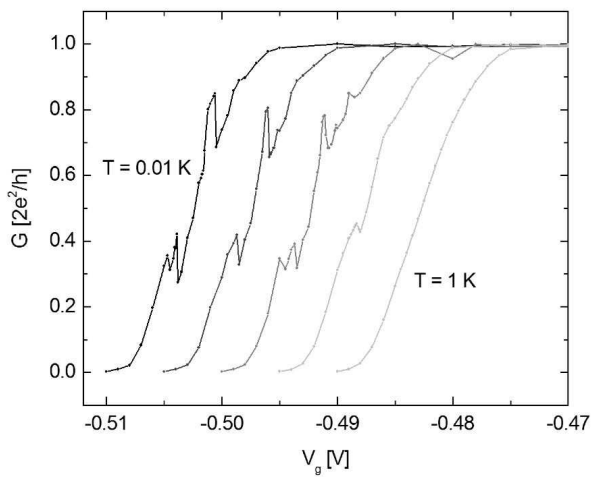}
\caption{Conductance versus gate voltage, for different
temperatures, 0.01 K, 0.05 K, 0.1 K, 0.5 K and 1 K. The traces are
shifted in voltage for clarity.} \label{fig:conduct}
\end{figure}

The two splittings give rise to two features in the conductance,
shown in Figure~\ref{fig:conduct}, at $V_{g}=-0.504$ V and
$V_{g}=-0.501$ V respectively.  Notably, the feature at
$V_{g}=-0.501$ V is strongly reminiscent of the $0.7G_{0}$
structure seen in numerous experiments. There also a  feature at
$0.4G_{0}$ seen simultaneously with the $0.7G_{0}$ in the absence
of external magnetic fields. Both features vanish with increasing
temperature, the $0.4G_{0}$ feature being slightly more robust
than the $0.7G_{0}$. We are not sure of the origins of the small
oscillations seen around the features at the lowest temperatures, but the general behaviour is robust \cite{comment}.

\section{Discussion}

The results presented above establish the relative insensitivity
of the 0.7 feature to temperature, since it persists up to 0.5 K, for the model we considered.
The reasons for the feature are largely compatible with
phenomenological models~\cite{kris,bruu,reil}, with a few
significant differences. In phenomenological model of \cite{reil}, the
subband energy of one spin configuration dips below the chemical
potential, whilst the other closely tracks the chemical potential
over some finite range of $V_{g}$.  In contrast, our results
indicate that the spin subbands do not split as they cross the
chemical potential, but rather after they have fallen below it.
After splitting, one subband (labelled ``Down'' in Figure \ref{fig:subbands})  rises through the chemical
potential, thereby suppressing its transmission coefficient around $V_g=-0.501 V$.

In contrast to many other theoretical and phenomenological
investigations of this subject, we have computed the transmission
coefficient from the self-consistent scattering potential. We
found that it varies slowly on the scale of any temperature we
consider, rather than behaving as a step function at the chemical
potential.  Furthermore, we are interested in the transmission
just below the first plateau, which is when the WKB method breaks
down. Thus, our treatment makes a fair estimate of $T(E)$.

Whilst the feature at $0.4 G_{0}$ was not entirely unexpected,
having been discussed in~\cite{ber2}, it is not commonly observed
in experiment. Recent observations show a structure at
around $0.4$ to $0.5 G_{0}$~\cite{crook}, which is ascribed to the mirror
symmetry of the channel, consistent with the calculation presented
here. This suggests a future avenue of exploration on the role of
asymmetry in the channel.  This is beyond the scope of the present
work. Nevertheless, the $0.4 G_{0}$ feature arises in our
calculation for the same reason as the $0.7 G_{0}$ feature: the
spin subbands split above the chemical potential, causing one of
them to approach the chemical potential over a narrow range of
$V_{g}$.

In conclusion, we have computed the self-consistent scattering
potential for a QPC at different temperatures, from which we
calculated the transmission coefficient and conductance. We found
features below the first conductance plateau consistent with the
experimentally observed $0.7 G_{0}$ structure, as well as another
at $0.4 G_{0}$. The features are robust against temperature, as
seen in experiment, but the high temperature behaviour that we
find is not in agreement with observation.

\section*{Appendix: Scattering Calculation}
\label{ap:scat}

Consider  a 1D potential $V(x)$ localised in the region $|x|<X$.
We compute the tunnelling amplitude for a particle entering the
region from the left with energy $E$.  The wavefunction takes the
form
 \begin{equation}
 \psi_E(x)=\left\{
 \begin{array}{ll}
      A e^{i k x}+B e^{-i k x}&  \textrm{for } x<-X \\
      \phi(x) &   \textrm{for } |x|<X\\
      C e^{i k x}& \textrm{for }x>X
\end{array}
\right.
\end{equation}
where $E=\hbar k_E^2/2m$, and both $\psi(x)$ and $\psi'(x)$ are
continuous everywhere.  The tunnelling probability is then given
by $T(E)=|C/A|^2$.  $\phi(x)$ satisfies the 1D Schr\"odinger
equation in the scattering region,
\begin{equation}
-\hbar^2/2m\, \phi''(x)+V(x)\phi(x)=E\, \phi(x).\label{eq:phi}
\end{equation}
To satisfy the matching conditions at $x=\pm X$, we first solve
\eqn{eq:phi} twice, using two linearly independent sets of
boundary conditions, such as $\{\phi_1(-X)=1,\phi_1(X)=0\}$ and
$\{\phi_2(-X)=0, \phi_2(X)=1\}$, leading to two linearly
independent solutions $\phi_1$ and $\phi_2$. These are then used
to construct $\phi(x)=\alpha \phi_1(x)+\beta \phi_2(x) $, and the
parameters $B, C, \alpha$ and $\beta$ are solved to satisfy the
four matching conditions.  With the boundary conditions above, and
for a symmetric scattering potential (so that
$\phi_2(x)=\phi_1(-x)$), we find
%\begin{equation}
%\frac{C}{A}=-\frac{2i e^{-2i k l}k \phi_1'(X)}{(k-i
%\phi_1'(-l))^2+\phi_1'(X)^2}.
%\end{equation}
\begin{equation}
\frac{C}{A}=-\frac{2i e^{-2i k X}k \phi_1'(X)}{(k-i
\phi_1'(-X))^2+\phi_1'(X)^2}.
\end{equation}

\section*{Acknowledgements}

We would like to thank D. J. Reilly, K.-F. Berggren, I. I.
Yakimenko, and A. C. Graham for fruitful discussions. This
research is part of the QIP IRC (www.qipirc.org) supported by
EPSRC (GR/S82176/01).


\begin{center}
\bf REFERENCES
\end{center}
\begin{enumerate}

\bibitem{whar}D. A. Wharam, T. J. Thornton, R. Newbury, M. Pepper, H. Ahmed, J. E. F. Frost,
D. G. Hasko, D. C. Peacock, D. A. Ritchie, and G. A. C. Jones, J.
Phys. C \textbf{21}, L209 (1988).

\bibitem{wees}B. J. van Wees, H. van Houten, C. W. J. Beenakker, J. G. Williamson,
L. P. Kouwenhoven, D. van der Marel, and C. T. Foxon, Phys. Rev.
Lett. \textbf{60}, 848 (1988).

\bibitem{but}M. B\"uttiker, Phys. Rev. B \textbf{41}, 7906 (1990).

\bibitem{fiel}M. Field, C. G. Smith, M. Pepper, D. A. Ritchie, J. E. F. Frost,
G. A. C. Jones, and D. G. Hasko, Phys. Rev. Lett. \textbf{70},
1311 (1993).

\bibitem{vand}L. M. K. Vandersypen, J. M. Elzerman, R. N. Schouten, L. H. Willems van Beveren,
R. Hanson, and L. P. Kouwenhoven, Appl. Phys. Lett. \textbf{85},
4394 (2004).

\bibitem{thom}K. J. Thomas, J. T. Nicholls, M. Y. Simmons, M. Pepper, D. R. Mace,
and D. A. Ritchie, Phys. Rev. Lett. \textbf{77}, 135 (1996).

\bibitem{fitz}R. Fitzgerald, Phys. Today \textbf{55} (5), 21 (2002).

\bibitem{ber}
C.-K. Wang and K.-F. Berggren, Phys. Rev. B {\bf 54}, R14 257
(1996); C.-K. Wang and K.-F. Berggren, Phys. Rev. B {\bf 57}, 4552
(1998).

\bibitem{bych}A. M. Bychkov, I. I. Yakimenko, and K.-F. Berggren,
Nanotechnology \textbf{11}, 318 (2000).

\bibitem{hiro}K. Hirose, S.-S. Li, and N. S. Wingreen, Phys. Rev.
B {\bf 63}, 033315 (2001).

\bibitem{ber2}
K.-F. Berggren and I. I. Yakimenko, Phys. Rev. B {\bf 66}, 085323
(2002).

\bibitem{hav} P. Havu, M. J. Puska, R. M. Nieminen and V. Havu, Phys. Rev. B {\bf 70}, 233308 (2004)

\bibitem{cron}S. M. Cronenwett, H. J. Lynch, D. Goldhaber-Gordon, L. P.
Kouwenhoven, C. M. Marcus, K. Hirose, N. S. Wingreen, and V.
Umansky, Phys. Rev. Lett. {\bf 88}, 226805 (2002).

\bibitem{meir}Y. Meir, K. Hirose, and N. S. Wingreen, Phys. Rev. Lett. {\bf 89},
196802 (2002); K. Hirose, Y. Meir, and N. S. Wingreen, Phys. Rev.
Lett. {\bf 90}, 026804 (2003).

\bibitem{kris}A. Kristensen, H. Bruus, A. E. Hansen, J. B. Jensen,
P. E. Lindelof, C. J. Marckmann, J. Nygard, C. B. Sorensen, F.
Beuscher, A. Forchel, and M. Michel, Phys. Rev. B {\bf 62}, 10950
(2000).

\bibitem{bruu}H. Bruus, V. V. Cheianov, K. Flensberg, Physica E {\bf 10},
97 (2001).

\bibitem{reil}D. J. Reilly, T. M. Buehler, J. L. O'Brien, A. R. Hamilton,
A. S. Dzurak, R. G. Clark, B. E. Kane, L. N. Pfeiffer, and K. W.
West, Phys. Rev. Lett. {\bf 89}, 246801 (2002); D. J. Reilly,
Phys. Rev. B {\bf 72}, 033309 (2005).

%\bibitem{crook}R. Crook, J. Prance, K. J. Thomas, I. Farrer, D. A.
%Ritchie, C. G. Smith, and M. Pepper, \textit{Spontaneous spin
%polarisation in a symmetric GaAs quantum wire}, (unpublished).

\bibitem{crook}R. Crook, J. Prance, K. J. Thomas, I. Farrer, D. A.
Ritchie, C. G. Smith, and M. Pepper, %\textit{Conductance Quantization at a Half-Integer Plateau in a Symmetric GaAs Quantum Wire}, 
Science {\bf 312}, 1359 (2006).

\bibitem{yak}
I. I. Yakimenko, A. M. Bychkov, and K.-F. Berggren, Phys. Rev. B
{\bf 63}, 165309 (2001).

\bibitem{PhD}A. M. Bychkov, D.Phil. thesis, Oxford University
(2003), available at http://cam.qubit.org/users/andrey/PhD.pdf.

\bibitem{tan}B. Tanatar and D. M. Ceperley, Phys. Rev. B \textbf{39}, 5005 (1989).

\bibitem{nano}K.-F. Berggren, I. I. Yakimenko, and A. M. Bychkov,
Nanotechnology \textbf{12}, 529 (2001).

\bibitem{star}
A. A. Starikov, I. I. Yakimenko, and K.-F. Berggren, Phys. Rev. B
{\bf 67}, 235319 (2003).

\bibitem{comment}
Our self-consistent process shows very slow convergence (requiring
up to 12000 iterations) at small but finite temperatures, for some values of
the applied gate voltage. Numerical instabilities have also been
reported for spin-density-functional studies of quantum wires in
Refs.~\cite{hiro,star}.

%\bibitem{mag} The asymmetry in the conductance as a function of the source-drain bias voltage of \cite{crook} suggests the existence of a local magnetic dipole, since the forward and reverse conductance should be the same in the absence of a magnetic fieldsince magnetic fields are not time-reversal invariant, 

\end{enumerate}
\end{document}